\documentclass[preprint]{aastex}
\usepackage{emulateapj5}
\slugcomment{To appear in the July 2002 issue of The Astronomical Journal}
\shorttitle{Planets in the Hyades I.}
\shortauthors{Cochran et al.}
\begin{document}

\title{Searching for Planets in the Hyades. \\
	I. The Keck Radial Velocity Survey \altaffilmark{1}
}
\altaffiltext{1}{Data presented herein were obtained
     at the W.M. Keck Observatory, which is operated as a
     scientific partnership among the California Institute of
     Technology, the University of California and the National
     Aeronautics and Space Administration. The Observatory was
     made possible by the generous financial support of the W.M.
     Keck Foundation.}

\author{William D. Cochran}
\affil{McDonald Observatory, The University of Texas at Austin,
       Austin, TX 78712}
\email{wdc@astro.as.utexas.edu}

\author{Artie P. Hatzes\altaffilmark{2}}
\affil{Th\"uringer Landessternwarte, D - 07778 Tautenburg, Germany}
\email{artie@jupiter.tls-tautenburg.de}

\and

\author{Diane B. Paulson}
\affil{Astronomy Department, The University of Texas at Austin,
       Austin, TX 78712}
\email{apodis@astro.as.utexas.edu}

\altaffiltext{2}{McDonald Observatory, The University of Texas at Austin,
       Austin, TX 78712}

\begin{abstract}
We describe a high-precision radial velocity search for jovian-mass companions
to main sequence stars in the Hyades star cluster.  The Hyades provides an
extremely well controlled sample of stars of the same age, the same
metallicity, and a common birth and early dynamical environment.  This sample
allows us to explore the dependence of the process of planet formation on only
a single independent variable: the stellar mass.  In this paper
we describe the survey and summarize results for the first five years.
\end{abstract}


\section{Introduction}
\label{intro}
The discoveries of jovian-mass companions around other stars are finally
supplying the observational data to test the widely held belief that
planet-formation should be a natural result of star-formation
\citep{MaCoMa00}.
The sub-stellar mass companions found so far demonstrate an astonishingly
wide diversity of systems.
There are jovian mass planets in orbits with semimajor axes as small as
10 stellar radii.  
Eccentric orbits appear to be the rule rather than the exception.
For practical purposes, most surveys so far have targeted relatively nearby
dwarf field stars similar to the Sun.
These inhomogeneous samples were adequate for discovering the first
extra-solar planets.
However, as planet search programs now work on the problem of 
characterization rather than mere discovery of extra-solar planetary systems,
the target lists must be much more carefully defined and selected in order to
limit the number of free parameters in the sample.
Several different types of surveys need to be undertaken, each
of which is designed to answer a specific focused scientific question
about the physics of planetary system formation.

We are conducting a survey of dwarf stars in the nearby Hyades star cluster
to investigate the dependence of planetary system formation on the mass of
the parent star.
Young open clusters have a number of advantages, as well as disadvantages as
samples for extrasolar planet surveys.
The target stars form a highly homogeneous sample of stars that is coeval
with a well determined age. 
The stars have uniform initial internal chemical composition,
removing issues of stellar metallicity.  The birth environment of the cluster
is relatively well understood.  The major independent variable in the sample
is the mass of the star.
While such a study can be conducted with field stars, one must be very careful
because field stars are not a homogeneous sample.   Their metallicities
can span a very wide range, the ages are difficult to determine well,
and their birth environments and locations are largely unknown.
Disentangling the complex intertwined effects of all of the factors that
might influence the properties of any detected planetary systems is
significantly more difficult
in a volume or magnitude limited sample of solar-neighborhood field stars.

In Section~\ref{hyades} we discuss the suitability of the Hyades as a
laboratory for exploring the dependence of planet formation on stellar mass.
Section~\ref{kecksurvey} presents the technical aspects of the radial
velocity survey using the Keck~1 telescope and its HIRES spectrometer.  
Preliminary results are presented in Section~\ref{results}.

\section{Searching for Planets in a Star Cluster}
\label{hyades}

The distribution of stellar masses in a star cluster is described by the 
the stellar initial mass function (IMF).
While the detailed physics that determines the actual shape of the IMF in a
cluster is still not well understood, it has become clear that there are not
huge variations in the IMF from one region to another \citep{MeAdHi00}.
We wish to investigate how, in turn, the mass of the star affects the properties
of any planetary system that may form around it.
Do massive stars form massive planets, or does tidal truncation \citep{LiPa86}
limit the size to which a planet can grow?
What role does early dynamical evolution play in systems with various mass
stars?
Do low mass stars form fewer planets, or perhaps lower mass planets?
The Hyades star cluster is an excellent place to seek answers to these
questions.

\subsection{Planet Formation in a Cluster Environment}
Stars in a cluster share a common birth environment, whereas the location and
early dynamical environment of field stars is largely unknown, except perhaps
for the sun \citep{AdLa01}.
The stellar birth environment can significantly influence the formation and
early evolution of both planetary and binary star systems. 
The planet-forming disks can be disrupted by close stellar encounters, and by
the UV  radiation from nearby hot stars.
Newly formed planetary systems are subject to tidal disruption in the dense
stellar environment of a young cluster.

\citet{Ar00} considered the effects of disk photoevaporation
by ultraviolet radiation from massive stars.  Giant planet formation can be
suppressed if the disk is destroyed on a timescale shorter than the planetary
core growth time of several million years.  Armitage computed the disk
photoevaporation lifetime as a function of the number of stars in the cluster
and the position of the star within the cluster, and concluded that this
process suppresses giant planet formation in clusters of $\sim10^5$ stars out
to a distance of about 1 parsec, and possibly farther from the cluster center.
If we accept the conventional wisdom that the very short-period giant planets
(e.g. 51~Peg\,b) are formed in the disk at several AU and then migrate
inward through tidal interactions with the disk \citep{BoHuLi00},
then the incidence of such short-period giant planet systems will be lower
in richer clusters because the disk must last sufficiently long
after planet formation to allow for the inward migration.
If, on the other hand, giant planets form by direct hydrodynamic collapse of
gravitational instabilities in the disk \citep{Bo00,Bo01}, then disk
photoevaporation is unlikely to affect the giant planet formation process
significantly.

Even if a planet-forming circumstellar disk is able to survive destruction by
UV photoevaporation processes, the system must still survive the effects of
close stellar encounters in the crowded cluster environment.
\citet{BoWaBh98} and \citet{WaBhBo98a,WaBhBo98b} investigated the effects of
star-disk encounters as well as disk-disk encounters (both coplanar and
non-coplanar).
These SPH code studies indicate that encounters may cause gravitational
instabilities which then lead to fragmentation of the disk and to the formation
of additional companion objects to the central star.   These simulations
assume a disk radius of 1000\,AU and model periastron encounter distances
ranging from 0.5 to 2.0 disk radii.
\citet{ScCl01} show that in a dense star forming
region such as the Orion Nebular Cluster (ONC), the distribution of stellar
encounter distances peaks at 1000\,AU, and only 10\% of the stars suffer an
encounter at less than 100\,AU in $10^7$~years.
The observed disk sizes in ONC tend to be significantly smaller than 1000\,AU
\citep{MCOD96} and thus Scally \& Clarke conclude that protoplanetary disks in
ONC are unlikely to be destroyed by close stellar encounters.  These results
are supported by computations by \citet{BoSmDa01} and by
\citet{SmBo01}, who find that stellar encounters do
not significantly affect planet formation in open clusters.

Even if a open cluster environment will not easily inhibit the formation of
giant planets, is it possible for stellar encounters to affect the planetary
system dynamical properties? 
\citet{FMFM97} proposed stellar encounters in open clusters as a means
of forming eccentric giant planets such as 16~Cygni~Bb \citep{CoHaBu97}.
They found that a small fraction of the giant planets formed in open clusters
could develop large eccentricities, but most systems will be unaffected.
\citet{LaAd98} found that while it is possible for
scattering by gravitational interactions with binary stars in the birth cluster
to account for some number of systems such as 16~Cygni~Bb and even for some
close-in giant planets such as 51~Peg\,b, the overall efficiency of the process
is too small to account for the observed semi-major axis -- eccentricity
distribution.

The stellar environment in a globular cluster, however, is significantly
different than in an open cluster. 
\citet{GiBrGu00} used HST WF/PC2 and STIS to search for photometric transits
of short period planets in 47~Tucanae.  
If main sequence stars in 47~Tuc have the same fraction
of 51~Peg-like systems as in the solar neighborhood, then this HST program
should have found about 15-20 transiting systems among the 34,000 stars
monitored for 8 consecutive days. 
Instead they found none, indicating that the formation rate of these
systems is significantly lower in this globular cluster. 
\citet{DaSi01} computed that during the evolution of 47~Tuc, 
wide planetary systems ($a \gtrsim 0.3$AU)
would probably be disrupted but tighter systems would probably survive,
especially in the less dense outer regions of the cluster.

\subsection{The Hyades as a Laboratory for Studying Planet Formation}
The Hyades are the closest and one of the most intensely studied open clusters.
The space motion of the Hyades has been extremely well determined
\citep{PeBrLe98,dBHodZ01} so cluster membership can be determined with high
confidence. 
The cluster currently comprises over 300 probable members, and a
total mass of 300-400\,M$_{\odot}$ \citep{PeBrLe98}. 
A large number of spectroscopic metallicity determinations of individual
Hyades stars \citep{ChCaSt71,BrLaTo80,CaCSCa85,BoBu88,Boesgaard89,BoFr90}
were combined by Perryman et~al. to give a mean metallicity of the Hyades of
$\mathrm{[Fe/H]} = +0.14 \pm 0.05$.
Much of the scatter in these  $\mathrm{[Fe/H]}$ values is likely due to
differences in the analysis techniques rather than real star-to-star abundance
variations. 
\citet{SmRu97} find a mean $\mathrm{[Fe/H]} = +0.13$ from analysis of
strong Ca~II 8542{\AA} and Mg~I 8806{\AA} lines. 
\citet{CaCSCa85} found that $\mathrm{[Fe/H]} = +0.12 \pm 0.03$ for
10~Hyades dwarfs, but that 2 other Hyades stars (HD\,27859 = vB\,52 and
HD\,27685 = vB\,39) had anomalously low
$\mathrm{[Fe/H]}$ of $+0.028$ which they attributed to the high level of
chromospheric activity in these stars.
\citet{PeBrLe98} fit a theoretical ZAMS to the Hyades, and derive a
helium content $Y = 0.26 \pm 0.02$, which agrees closely with the value
obtained by \citet{LeFeLe01} of $Y = 0.255 \pm 0.013$ as well as
with the solar value of $Y = 0.2659$.
Isochrone fitting to the Hyades main sequence turn off then gives a cluster ago
of $625 \pm 50$\,My, in excellent agreement with previous determinations of
655\,My \citep{CS90} and 600\,My \citep{ToStLa97}.
\citet{dBHodZ01} used the Hipparcos proper motions to
derive secular parallaxes for Hyades members which they claimed were
significantly better than the Hipparcos trigonometric parallaxes.  While this
analysis does indeed narrow the width of the Hyades main sequence, it
does not really alter the theoretical fit to the main sequence or the derived
values of $Y$ and $Z$.  However, the very narrow width of the main sequence
does support the lack of significant star-to-star metallicity variations among
Hyades dwarfs.

While the Hyades are a nearby, homogeneous sample of stars, are they really
suitable targets for high-precision radial velocity work?  Achieving high
velocity precision requires SNR of about 300 in a relatively short exposure on
stars that are slowly rotating and are relatively inactive.  Hyades dwarfs
range in magnitude from about $V = 7.5$ for a late F star through $V = 9 - 11$
for K stars.  The M dwarfs range down through significantly fainter magnitudes.
The necessary SNR can be achieved on these stars in exposures of 15 minutes
or less with the Keck HIRES spectrograph \citep{VoAlBi94}.
The Hyades dwarfs, at their age of 625\,My, have slowed their rotation rates
significantly from the rotation rates found in younger clusters such as the
Pleiades \citep{TeStPi00}.  Hyades dwarfs show a good correlation between
rotation rates and activity \citep{StBaKr97,TeStPi00}.
We have selected stars of low $v \sin i$ for our Hyades sample.
The major question that we face in the use of Hyades dwarfs for precise radial
velocity studies is whether the level of stellar activity in theses stars will
be low enough that we will still be able to detect RV variations due to orbital
reflex motion on top of the level of intrinsic RV jitter in these stars.
\citet{SaDo97}, \citet{SaBuMa98}, and \citet{SaMaNa00} have derived empirical
relationships between the observed radial velocity ``jitter'' of single stars
presumed not to have planetary mass companions and observable stellar
properties such as $v \sin i$ and R$^{\prime}_{HK}$.
The applicability of such relationships to our Hyades data set is the major
subject of Paper~II in this series \citep{PaSaCo02}.

\section{The Keck Radial Velocity Survey of Hyades Dwarfs}
\label{kecksurvey}
Since 1996 we have been using the Keck~1 telecope with its HIRES
\citep{VoAlBi94} spectrograph to conduct high precision radial velocity
observations of a sample of 98 Hyades dwarf stars.
An I$_2$ gas absorption cell provides the velocity metric
\citep{Li88,BuMaWi96}.
The HIRES spectrograph is set to include the stellar Ca~II H\,\&\,K lines
(3933.66, 3968.47{\AA}).  These lines fall in a wavelength region free of I$_2$
gas cell absorption, and thus provide a direct and simultaneous measurement of
stellar chromospheric activity for every radial velocity observation.
This automatic ability to monitor stellar activity has turned out to be
extremely important for the relatively young, chromospherically
active Hyades dwarfs.
The red limit of the spectrograph setting is 6188{\AA}, near
the effective red limit of useful I$_2$ gas absorption.

The HIRES ``B2'' entrance aperture ($0.574\arcsec \times 7.0\arcsec$
projected on
the sky) gives a nominal resolving power of 60,000 and an actual measured
resolving power ranging up to 67,000 depending on position on the CCD.
The image derotator is used to keep the slit vertical on the sky.
This keeps atmospheric dispersion perpendicular to the spectrograph echelle
dispersion, and presents a constant orientation of the telescope pupil to
the spectrograph.  The Keck HIRES exposure meter, installed in 2000,
is used to terminate exposures when the
desired SNR is reached, and to give an accurate determination of the
photon-weighted mid-exposure time.
We did not apply the empirical correction for problems in the HIRES CCD readout
electronics derived by \citet{VoMaBu00}.

The target stars were selected in 1996, before the Hipparcos data for the
Hyades stars were released.   After the analysis by 
\citet{PeBrLe98} our target list was significantly modified.
Several non-members were dropped, and a few new stars were added.
The observing list now comprises 98 Hyades dwarfs: 10 F~stars, 24 G~stars, 44
K~stars and 20 M~stars.
Since the object of this program is to explore the dependence of planet
formation on stellar mass, we have taken great care to include a
representative sample of lower mass stars. 
Due to the steepness of the stellar mass-luminosity
relation, the overwhelming majority of the telescope time is spent
observing these late K and M dwarfs. 
The stars have been selected to meet the following criteria:
1) classified as a Hyades member by \citet{PeBrLe98},
2) commonly accepted by other studies as a Hyades member based on radial
velocity, parallax, and proper motion, 3) not in a spectroscopic binary
system with orbital period less than $\sim 100$ years, and
4) stellar rotational velocity $v \sin i \leq 15$\,km\,s$^{-1}$.  
The motivation for criteria 1 and 2 is obvious; this is what guarantees
the homogeneity of our sample.
The prohibition on spectroscopic binary stars is a common feature of virtually
all RV planet detection programs.  It simply guarantees that there will not be
a stellar companion in an orbit small enough to prevent the presence of
planetary companions.  The limit to small $v \sin i$ is also a common feature
of high precision RV programs, and is predicated on practical considerations.
Narrow stellar lines are necessary to achieve the very high radial velocity
precision required to detect planetary companions.
This is the one selection criterion which might introduce some small
biases in our stellar sample. 
Low rotational velocities will probably select for stars with
lower levels of stellar activity \citep{StBaKr97}. 
In addition, if the stellar rotation rate at the 625\,My
age of the Hyades is related to the degree of rotational braking of the star by
magnetic coupling to a remnant circumstellar disk at earlier times, we may be
selecting for stars that had more massive disks, or stars for which the disks
survived longer.   However, in practice, our $v \sin i$  criterion was 
actually more of a spectral class criterion.  We rejected eight F5 stars, three
F6 stars, and one G5 star based solely on $v \sin i$.  Of the ten F stars
remaining in the survey, we have one F5 star, eight F8 stars, and one F9 star.
Thus, a possible bias toward low $\sin i$ is probably applicable only to
the sole surviving F5 star HD\,29419 (vB\,105).
We will live with this possible selection
effect and bear it in mind when interpreting the results of our survey.

\begin{figure}
\plotone{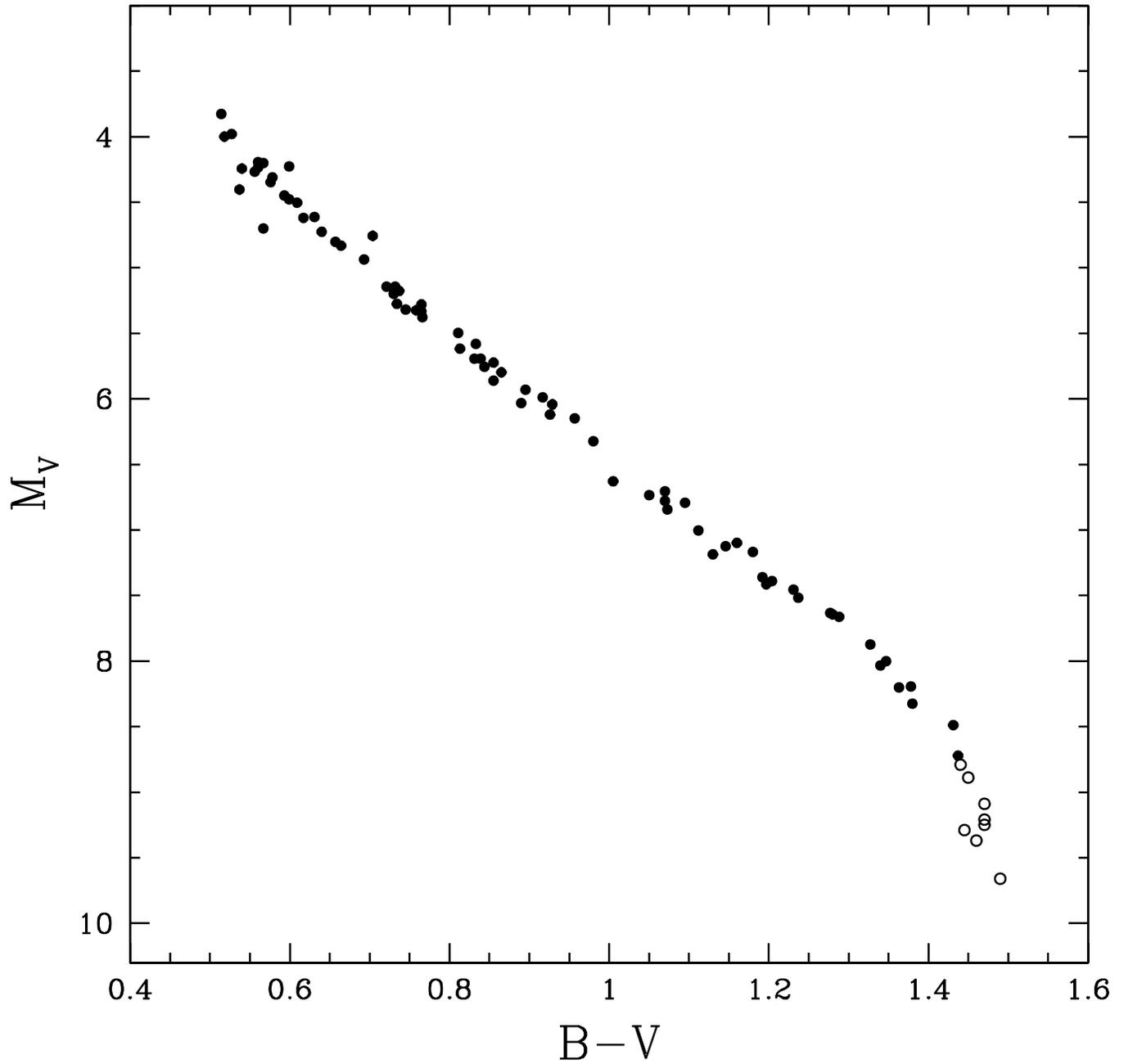}
\caption{Color-Magnitude diagram for Hyades program stars.  The solid circles
are program stars in the Hipparcos catalog.  The absolute magnitudes for these
stars have been computed using the statistical parallaxes from de~Bruijne
{\itshape et~al.} (2001).   The open circles are stars too faint to be included
in the Hipparcos survey.  For these stars, absolute magnitudes from van~Altena
(1966) were used.}
\label{cmdiag}
\end{figure}

A color-magnitude diagram of our program stars for which parallaxes are
available is given in Figure~\ref{cmdiag}.
The absolute magnitudes for the stars are computed using
the secular parallaxes of de~Bruijne et~al. (2001)  where ever
possible.  For those stars too faint to have been included in the Hipparcos
catalog, absolute magnitudes from \citet{vA66} were used.

\section{Preliminary Results}
\label{results}
This observing program has received an average of five to six nights per year
on the Keck~1 telescope, starting in the fall of 1996.
Since these target stars are on the average about three magnitudes fainter than
the targets for most other  high precision radial velocity surveys, our Hyades
survey has been hard-pressed to get good temporal coverage of all of the target
stars.  We have tried to obtain at least two good quality velocity measurements
of each star per year.  The sampling obtained by our scheduling on Keck has
been very good for sampling long periods, but extremely poor for sampling
possible short periods variations.

\begin{figure}
\epsscale{0.85}
\plotone{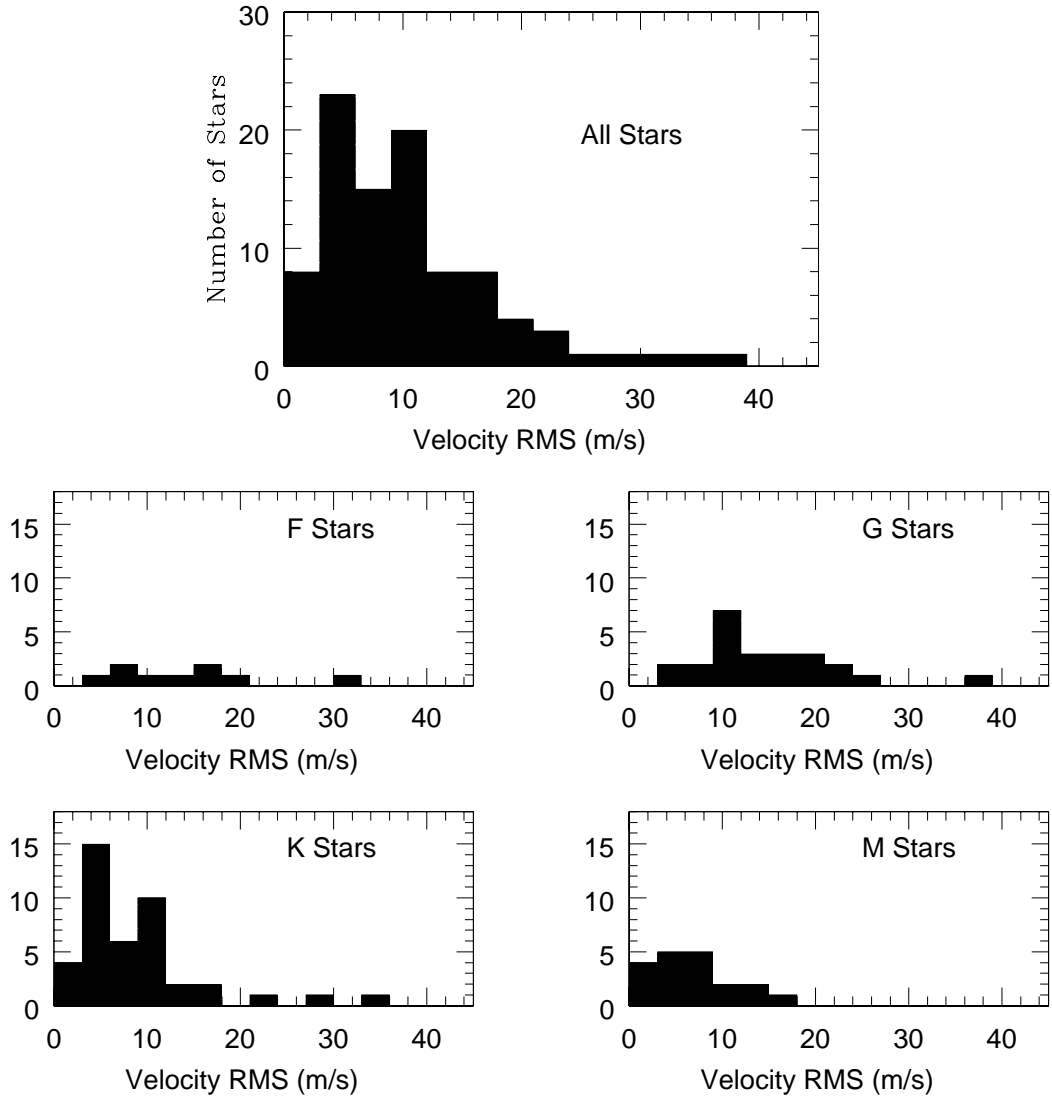}
\caption{The distribution of variances of the velocity measurements of program
stars about the mean, after removal of any long-term trends.  The upper panel
gives results for all stars in the program, and the lower panels break out the
results by stellar spectral class.   The histograms do not include four stars
(1 F star, 2 K stars, and 1 M star) with velocity RMS greater than
40\,m\,s$^{-1}$.}
\label{rms}
\end{figure}

We have attempted to assess the velocity precision we have achieved with
Keck/HIRES.  Figure~\ref{rms} gives histograms of the variance of the velocity
measurements about the mean for our program stars, after any linear trend has
been removed.   These observed variances will be a combination of measurement
uncertainty, intrinsic stellar variability, and possible orbital motion.
The median observed rms for the entire sample of Hyades dwarfs is
9.3\,m\,s$^{-1}$.  The lower panels of Figure~\ref{rms} give the distribution
of observed variances for the different spectral classes in the sample.  We see
that in general the observed variance is larger for the F and G stars than it
is for the K and M stars.  The variances for the stars at the low end of each
distribution is undoubtedly dominated by the uncertainty in the measurement
process, while the higher variances are probably real.
Our velocity measurement procedure gives us an independent assessment of the
internal velocity precision which is most likely to be dominated by intrinsic
measurement uncertainties.  In measuring the velocities for each observation,
we divide the spectrum into several hundred small intervals, and perform the
spectral modeling for each spectral interval independently. 
The variance of the velocities computed for each spectral interval is then
used to compute an error bar for the mean velocity of each observation.  
While we normally refer to this as an
``internal'' error, it is possible that some portion could be intrinsic to the
star.  For example, if there are velocity fields within the stellar
photosphere, then weak photospheric lines could easily have a very different
velocity behavior than strong lines.   This velocity difference would also vary
with any stellar activity cycle.  Nevertheless, the mean ``internal'' error for
the F stars is 6.8\,m\,s$^{-1}$, 4.7\,m\,s$^{-1}$ for the G stars,
4.3\,m\,s$^{-1}$ for the K stars, and 5.2\,m\,s$^{-1}$ for the M stars.
The observed trend in internal errors with spectral type is the result of the
interplay of two different factors: photon statistics, and the intrinsic
velocity content of the spectrum.  Since our Hyades stars are all at roughly
the same distance, the apparent magnitude of a star is a simple monotonic
function of the spectral type. 
Since we have imposed a maximum exposure time of
15~minutes on all of our observations in order to minimize uncertainties in the
barycentric correction, observations of the later spectral type stars have
lower SNR than the earlier spectral types.   Thus, we expect photon statistics
to become increasingly important for later spectral types.   The competing
effect is simply that the stellar spectrum becomes more complex with decreasing
stellar effective temperature.  The intrinsic velocity information content of a
stellar spectrum depends on the mean absolute value of the slope of the
spectrum \citep{BuMaWi96}.  This increases significantly for the later spectral
types \citep{BoPeQu01}.   Thus, we are in the fortunate position of having
roughly constant internal measurement errors of 4-6\,m\,s$^{-1}$ for all of our
Hyades sample, independent of stellar mass.   Paper~II discusses the
relationship between our measured radial velocity variations of Hyades dwarfs,
and stellar chromospheric activity.   The fundamental conclusion is that we are
able to achieve sufficient velocity precision in our sample of Hyades dwarfs,
in spite of their relatively young age, to detect jovian mass planetary
companions in orbit around them.   Long timescale stellar activity variations,
while present and significant, does not prevent us from achieving the
fundamental goals of our observing program.

\begin{figure}
\epsscale{0.80}
\plotone{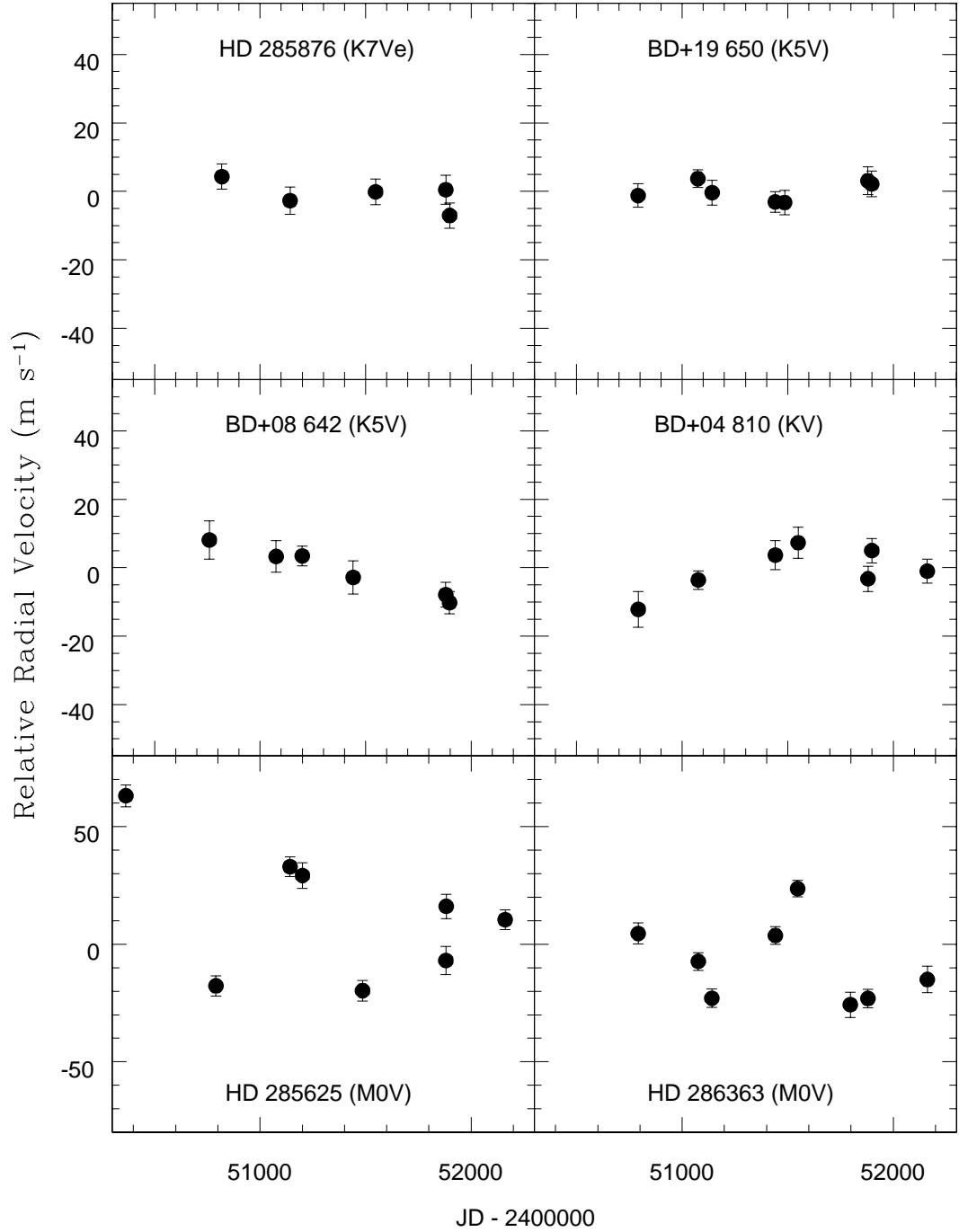}
\caption{Observed velocities for six typical stars in our Keck Hyades program.
The upper two panels show two stars constant to 3\,m\,s$^{-1}$ over 4 years
of observation.  The middle panels show two stars with secular velocity trends.
The lower two panels show stars with large variances which are not explained by
the error bars of the individual measurements.   We believe that the observed
variations are intrinsic to the stellar system in some manner.  These will be
followed up using the Hobby{\Large$\cdot$}Eberly Telescope.}
\label{stars}
\end{figure}

Figure~\ref{stars} shows examples of the velocity measurements we have
obtained.  The upper two panels show two stars, HD~285876 (HIP~21138, vB~191)
and BD+19~650 (HIP~18946) which show velocities over four years which
are constant to 3\,m\,s$^{-1}$.
The middle two panels show two stars, BD+08~642 (HIP~19441) and
BD+04~810 (HIP~23312) with secular trends in the data,
but with an rms about that trend of 3-4\,m\,s$^{-1}$.
The lower two panels show two stars with large scatter in their observed
velocities.  HD~285625 (HIP~19834) shows a velocity rms of 53\,m\,s$^{-1}$
and HD~286363 (HIP~18322) shows a velocity rms of 16\,m\,s$^{-1}$. 
Both of these last stars are candidates for having short period planetary
companions. 
Definitive detection of short period velocity variability has proven to
be extremely difficult with the scheduling of this program on Keck. 
Each star is typically observed only twice per year, leading to significant
period aliasing and ambiguity for short period RV variations. 
To alleviate this problem, we are performing follow-up
observations of stars showing significant RV scatter using the high resolution
spectrograph (HRS) of the Hobby{\Large$\cdot$}Eberly Telescope (HET).
The queue scheduled nature of the HET will allow us to optimally sample a range
of possible short periods.  At the conclusion of this Hyades survey, we will
place upper limits on companion $m \sin i$ as a function of orbital period
for each of our survey targets and we will then compare the frequency of
planetary companions to Hyades stars with similar results for field stars.

\section{Conclusions}
\label{conclusions}
Star clusters, such as the Hyades, provide an excellent laboratory for
investigating the physics of planetary system formation.  Most radial velocity
surveys for extrasolar planets have used samples of stars with widely ranging
and often unknown ages, metallicities, birth environments, and evolutionary
histories.  However, in the Hyades we have a very well controlled and
understood sample of stars in which the primary independent variable is the
stellar mass.  The goal of this program is to attempt to understand the
dependence of the properties of planetary systems on the mass of the central
star.  With the Hyades we have demonstrated 3--6\,m\,s$^{-1}$ radial velocity
precision for main sequence stars between F8 and M2, using the Keck~1 telescope
and its HIRES spectrograph with an I$_2$ cell as the velocity metric.   While 
these stars are indeed chromospherically active, the activity is not large
enough to seriously interfere with our ability to detect jovian-mass
companions.   A number of promising candidate planetary systems have been
identified which will be followed using the Hobby{\Large$\cdot$}Eberly
Telescope.

\acknowledgments
This work was supported by NASA grants NAG5-4384 and NAG5-9277, and by NSF
grant AST-9808980.   We wish to thank Geoff Marcy, Paul Butler and Steve Vogt
for valuable advice in the use of Keck HIRES, and for permission to use the
HIRES I$_2$ cell in advance of its public release. 
We are extremely grateful to David Latham and Robert Stefanik
for their assistance in removing binary stars and Hyades non-members
from our star sample, and to Debra Fischer for several helpful comments
and suggestions.

\begin{thebibliography}{44}
\expandafter\ifx\csname natexlab\endcsname\relax\def\natexlab#1{#1}\fi

\bibitem[{Adams \& Laughlin(2001)}]{AdLa01}
Adams, F.~C. \& Laughlin, G. 2001, Icarus, 150, 151

\bibitem[{Armitage(2000)}]{Ar00}
Armitage, P.~J. 2000, A\&A, 362, 968

\bibitem[{Bodenheimer {et~al.}(2000)Bodenheimer, Hubickyi, \&
  Lissauer}]{BoHuLi00}
Bodenheimer, P., Hubickyi, O., \& Lissauer, J. 2000, Icarus, 143, 2

\bibitem[{Boesgaard(1989)}]{Boesgaard89}
Boesgaard, A.~M. 1989, ApJ, 336, 798

\bibitem[{Boesgaard \& Budge(1988)}]{BoBu88}
Boesgaard, A.~M. \& Budge, K.~G. 1988, ApJ, 332, 410

\bibitem[{Boesgaard \& Friel(1990)}]{BoFr90}
Boesgaard, A.~M. \& Friel, E.~D. 1990, ApJ, 351, 467

\bibitem[{Boffin {et~al.}(1998)Boffin, Watkins, Bhattal, Francis, \&
  Whitworth}]{BoWaBh98}
Boffin, H. M.~J., Watkins, S.~J., Bhattal, A.~S., Francis, N., \& Whitworth,
  A.~P. 1998, MNRAS, 300, 1189

\bibitem[{Bonnell {et~al.}(2001)Bonnell, Smith, Davies, \& Horne}]{BoSmDa01}
Bonnell, I.~A., Smith, K.~W., Davies, M.~D., \& Horne, K. 2001, MNRAS, 322, 859

\bibitem[{Boss(2000)}]{Bo00}
Boss, A.~P. 2000, ApJ, 536, L101

\bibitem[{Boss(2001)}]{Bo01}
---. 2001, ApJ, 551, L167

\bibitem[{Bouchy {et~al.}(2001)Bouchy, Pepe, \& Queloz}]{BoPeQu01}
Bouchy, F., Pepe, F., \& Queloz, D. 2001, A\&A, 374, 733

\bibitem[{Branch {et~al.}(1980)Branch, Lambert, \& Tomkin}]{BrLaTo80}
Branch, D., Lambert, D.~L., \& Tomkin, J. 1980, ApJ, 241, L83

\bibitem[{Butler {et~al.}(1996)Butler, Marcy, Williams, McCarthy, Dosanjh, \&
  Vogt}]{BuMaWi96}
Butler, R.~P., Marcy, G.~W., Williams, E., McCarthy, C., Dosanjh, P., \& Vogt,
  S.~S. 1996, PASP, 108, 500

\bibitem[{Cayrel {et~al.}(1985)Cayrel, {Cayrel de Strobel}, \&
  Campbell}]{CaCSCa85}
Cayrel, R., {Cayrel de Strobel}, G., \& Campbell, B. 1985, A\&A, 145, 249

\bibitem[{{Cayrel de Strobel}(1990)}]{CS90}
{Cayrel de Strobel}, G. 1990, Memorie della {S}ocieta {A}stronomica {I}taliana,
  61, 613

\bibitem[{Chaffee {et~al.}(1971)Chaffee, Carbon, \& Strom}]{ChCaSt71}
Chaffee, F.~H., Carbon, D.~F., \& Strom, S.~E. 1971, ApJ, 166, 593

\bibitem[{Cochran {et~al.}(1997)Cochran, Hatzes, Butler, \& Marcy}]{CoHaBu97}
Cochran, W.~D., Hatzes, A.~P., Butler, R.~P., \& Marcy, G.~W. 1997, ApJ, 483,
  457

\bibitem[{Davies \& Sigurdsson(2001)}]{DaSi01}
Davies, M.~B. \& Sigurdsson, S. 2001, MNRAS, 324, 612

\bibitem[{{de Bruijne} {et~al.}(2001){de Bruijne}, Hoogerwerf, \& {de
  Zeeuw}}]{dBHodZ01}
{de Bruijne}, J. H.~J., Hoogerwerf, R., \& {de Zeeuw}, P.~T. 2001, A\&A, 367,
  111

\bibitem[{{de la Fuente Marcos} \& {de la Fuente Marcos}(1997)}]{FMFM97}
{de la Fuente Marcos}, C. \& {de la Fuente Marcos}, R. 1997, A\&A, 326, L21

\bibitem[{Gilliland {et~al.}(2000)Gilliland, Brown, Guhathakurta, Sarajedini,
  Milone, Albrow, Baliber, Bruntt, Burrows, Charbonneau, Choi, Cochran,
  Edmonds, Frandsen, Howell, Lin, Marcy, Mayor, Naef, Sigurdsson, Stagg,
  Vandenberg, Vogt, \& Williams}]{GiBrGu00}
Gilliland, R.~L., Brown, T.~M., Guhathakurta, P., Sarajedini, A., Milone,
  E.~F., Albrow, M.~D., Baliber, N.~R., Bruntt, H., Burrows, A., Charbonneau,
  D., Choi, P., Cochran, W.~D., Edmonds, P.~D., Frandsen, S., Howell, J.~H.,
  Lin, D. N.~C., Marcy, G.~W., Mayor, M., Naef, D., Sigurdsson, S., Stagg,
  C.~R., Vandenberg, D.~A., Vogt, S.~S., \& Williams, M.~D. 2000, ApJ, 545, L47

\bibitem[{Laughlin \& Adams(1998)}]{LaAd98}
Laughlin, G. \& Adams, F.~C. 1998, ApJ, 508, L171

\bibitem[{Lebreton {et~al.}(2001)Lebreton, Fernandes, \& Lejeune}]{LeFeLe01}
Lebreton, Y., Fernandes, J., \& Lejeune, T. 2001, A\&A, 374, 540

\bibitem[{Libbrecht(1988)}]{Li88}
Libbrecht, K.~G. 1988, in {I.A.U.} {S}ymposium 132: {T}he Impact of Very High
  {S/N} Spectroscopy on Stellar Physics, ed. G.~{Cayrel de Strobel} \& M.~Spite
  (Dordrecht: Kluwer), 83

\bibitem[{Lin \& Papaloizou(1986)}]{LiPa86}
Lin, D. N.~C. \& Papaloizou, J. 1986, ApJ, 307, 395

\bibitem[{Marcy {et~al.}(2000)Marcy, Cochran, \& Mayor}]{MaCoMa00}
Marcy, G.~W., Cochran, W.~D., \& Mayor, M. 2000, in Protostars and Planets IV,
  ed. V.~Mannings, A.~P. Boss, \& S.~S. Russell (University of Arizona Press),
  1285

\bibitem[{McCaughrean \& O'Dell(1996)}]{MCOD96}
McCaughrean, M.~J. \& O'Dell, C.~R. 1996, AJ, 111, 1977

\bibitem[{Meyer {et~al.}(2000)Meyer, Adams, Hillenbarnd, Carpenter, \&
  Larson}]{MeAdHi00}
Meyer, M.~R., Adams, F.~C., Hillenbarnd, L.~A., Carpenter, J.~M., \& Larson,
  R.~B. 2000, in Protostars and Planets IV, ed. V.~Mannings, A.~P. Boss, \&
  S.~S. Russell (University of Arizona Press), 121

\bibitem[{Paulson {et~al.}(2002)Paulson, Saar, Cochran, \& Hatzes}]{PaSaCo02}
Paulson, D.~B., Saar, S.~H., Cochran, W.~D., \& Hatzes, A.~P. 2002, submitted
  to {\itshape A. J.}

\bibitem[{Perryman {et~al.}(1998)Perryman, Brown, Lebreton, G\'omez, Turon,
  {Cayrel de Strobel}, Mermilliod, Robichon, Kovalevsky, \& Crifo}]{PeBrLe98}
Perryman, M. A.~C., Brown, A. G.~A., Lebreton, Y., G\'omez, A., Turon, C.,
  {Cayrel de Strobel}, G., Mermilliod, J.~C., Robichon, N., Kovalevsky, J., \&
  Crifo, F. 1998, A\&A, 331, 81

\bibitem[{Saar {et~al.}(1998)Saar, Butler, \& Marcy}]{SaBuMa98}
Saar, S.~H., Butler, R.~P., \& Marcy, G.~W. 1998, ApJ, 498, L153

\bibitem[{Saar \& Donahue(1997)}]{SaDo97}
Saar, S.~H. \& Donahue, R.~A. 1997, ApJ, 485, 319

\bibitem[{Santos {et~al.}(2000)Santos, Mayor, Naef, Pepe, Queloz, Udry, \&
  Blecha}]{SaMaNa00}
Santos, N.~C., Mayor, M., Naef, D., Pepe, F., Queloz, D., Udry, S., \& Blecha,
  A. 2000, A\&A, 361, 265

\bibitem[{Scally \& Clarke(2001)}]{ScCl01}
Scally, A. \& Clarke, C. 2001, MNRAS, 325, 449

\bibitem[{Smith \& Ruck(1997)}]{SmRu97}
Smith, G. \& Ruck, M.~J. 1997, A\&A, 324, 1091

\bibitem[{Smith \& Bonnell(2001)}]{SmBo01}
Smith, K.~W. \& Bonnell, I.~A. 2001, MNRAS, 322, L1

\bibitem[{Stauffer {et~al.}(1997)Stauffer, Balachandran, Krishnamurthi,
  Pinsonneault, Terndrup, \& Stern}]{StBaKr97}
Stauffer, J.~R., Balachandran, S.~C., Krishnamurthi, A., Pinsonneault, M.,
  Terndrup, D.~M., \& Stern, R.~A. 1997, ApJ, 475, 604

\bibitem[{Terndrup {et~al.}(2000)Terndrup, Stauffer, Pinsonneault, Sills, Yuan,
  Jones, Fischer, \& Krishnamurthi}]{TeStPi00}
Terndrup, D.~M., Stauffer, J.~R., Pinsonneault, M., Sills, A., Yuan, Y., Jones,
  B.~F., Fischer, D., \& Krishnamurthi, A. 2000, AJ, 119, 1303

\bibitem[{Torres {et~al.}(1997)Torres, Stefanik, \& Latham}]{ToStLa97}
Torres, G., Stefanik, R.~P., \& Latham, D.~W. 1997, ApJ, 474, 256

\bibitem[{{van Altena}(1966)}]{vA66}
{van Altena}, W.~F. 1966, AJ, 71, 482

\bibitem[{Vogt {et~al.}(1994)Vogt, Allen, Bigelow, Bresee, Brown, Cantrall,
  Conrad, Couture, Delaney, Epps, Hilyard, Hilyard, Horn, Jern, Kanto, Keane,
  Kibrick, Lewis, Osborne, Pardeilhan, Pfister, Ricketts, Robinson, Stover,
  Tucker, Ward, \& Wei}]{VoAlBi94}
Vogt, S.~S., Allen, S.~L., Bigelow, B.~C., Bresee, L., Brown, B., Cantrall, T.,
  Conrad, A., Couture, M., Delaney, C., Epps, H.~W., Hilyard, D., Hilyard,
  D.~F., Horn, E., Jern, N., Kanto, D., Keane, M.~J., Kibrick, R.~I., Lewis,
  J.~W., Osborne, J., Pardeilhan, G.~H., Pfister, T., Ricketts, T., Robinson,
  L.~B., Stover, R.~J., Tucker, D., Ward, J., \& Wei, M.~Z. 1994, Proc. Soc.
  Photo-opt. Inst. Eng., 2198, 362

\bibitem[{Vogt {et~al.}(2000)Vogt, Marcy, Butler, \& Apps}]{VoMaBu00}
Vogt, S.~S., Marcy, G.~W., Butler, R.~P., \& Apps, K. 2000, ApJ, 536, 902

\bibitem[{Watkins {et~al.}(1998{\natexlab{a}})Watkins, Bhattal, Boffin,
  Francis, \& Whitworth}]{WaBhBo98a}
Watkins, S.~J., Bhattal, A.~S., Boffin, H. M.~J., Francis, N., \& Whitworth,
  A.~P. 1998{\natexlab{a}}, MNRAS, 300, 1205

\bibitem[{Watkins {et~al.}(1998{\natexlab{b}})Watkins, Bhattal, Boffin,
  Francis, \& Whitworth}]{WaBhBo98b}
---. 1998{\natexlab{b}}, MNRAS, 300, 1214
\end{thebibliography}
\end{document}